# Substrate induced tuning of compressive strain and phonon modes in large area MoS$_2$ and WS$_2$ van der Waals epitaxial thin films

Rajib Sahu[1], Dhanya Radhakrishnan[2], Badri Vishal[1], Devendra Singh Negi[1], Anomitra. Sil[3], Chandrabhas Narayana[2], and Ranjan Datta[1,*]

[1]International Centre for Materials Science, Jawaharlal Nehru Centre for Advanced Scientific Research, Bangalore 560064, India.

[2] Light Scattering Laboratory, Chemistry and Physics of Materials Unit, Jawaharlal Nehru Centre for Advanced Scientific Research, Bangalore 560064, India.

[3] Center for Nano Science and Engineering, Indian Institute of Science, Bangalore 560012, India.

**ABSTRACT:** Large area MoS$_2$ and WS$_2$ van der Waals epitaxial thin films with complete control over number of layers including monolayer is grown by pulsed laser deposition utilizing slower growth kinetics. The films grown on *c*-plane sapphire show stiffening of A$_{1g}$ and E$^1_{2g}$ phonon modes with decreasing number of layers for both MoS$_2$ and WS$_2$. The observed stiffening translate into the compressive strain of 0.52 % & 0.53 % with accompanying increase in fundamental direct band gap to 1.74 and 1.68 eV for monolayer MoS$_2$ and WS$_2$, respectively. The strain decays with the number of layers. HRTEM imaging directly reveals the nature of atomic registry of van der Waals layers with the substrate and the associated compressive strain. The results demonstrate a practical route to stabilize and engineer strain for this class of material over large area device fabrication.

*Corresponding author email: ranjan@jncasr.ac.in

# 1. Introduction

Atomically thick two dimensional layer materials generated immense research interest since the discovery of graphene [1-6]. Graphene revealed many unique physical phenomena e.g., linear band structure with momentum, ballistic transport [7-10] and superior chemical catalytic activities etc. [11, 12]. However, graphene is a zero band gap material and the need to open a band gap for practical device application shifted the attention towards $MoS_2$, $WS_2$ and similar materials belonging to the family of layered transition metal dichalcogenides (TMDCs) [13, 14]. Most of them exhibit indirect to direct band gap crossover from bulk to monolayer form and promising for extremely thin transistors, exploring fundamental physics, and energy related applications. Most of the fundamental properties based on this system are investigated primarily on mechanically exfoliated and chemically synthesized materials [15-19] which are not suitable for large area practical device fabrication. There have been many attempts to grow such van der Waals materials epitaxially over large area and explore the tuning of electronic structure through the application of strain [16-27]. Most of the epitaxial layer growth reported was only in the nano or micrometer length scale along with special treatment of substrate and the application of strain was performed by external means. For example, large area CVD growth of $MoS_2$ thin films was reported on $SiO_2$ substrate [18, 28]. Control over the number of layers was achieved by the duration of oxygen plasma treatment of $SiO_2$ substrate. Oxygen plasma treatment was believed to increase the interaction between the $MoS_2$ and $SiO_2$ substrate and helps in further growth of subsequent layers [28]. The characteristics Raman mode are different for the thin films grown on the substrate compared to the bulk suggesting the substrate induced van der Waals and long range Coulomb interaction. Most of the reports show softening of the Raman modes upon decreasing the number of layers grown on the widely used $SiO_2$ substrate. Among various reports on the tuning the electronic structure by strain is one where externally applied uniaxial tensile strain on a fabricated $MoS_2$ bending device resulted in phonon mode softening. This phonon mode softening was associated with the decrease in band gap by ~45 meV and ~120 meV for every 1% of strain for monolayer and bilayers, respectively. Direct to indirect crossover occurs for monolayer MoS$_2$ with applied strain of 1% [29]. On the other hand, application of bi-axial compressive strain was shown to increase the band gap of MoS$_2$ monolayer and the change in band gap is within 100 meV for every 1% of strain [30]. Nevertheless, from practical point of view it is important to grow such layered compounds epitaxially over large area with complete control over the layer thickness and at the same time inducing strain through the substrate in order to retain the modification in the electronic structure. Both the aspects of large area (we have grown on maximum 8×8 mm$^2$ size substrate) epitaxial film growth and retaining strained state are demanding and challenging. With this goal, in the present report we first demonstrate that it is possible to grow epitaxial MoS$_2$ and WS$_2$ thin films by pulsed laser deposition (PLD) under slow kinetic condition on '$c$' plane sapphire substrate [31] which has some advantages over the CVD counterpart. Moreover, complete control over layer thickness is possible by this method including the growth of monolayer without requiring any special substrate treatment. An important second part of the results is that the sapphire substrate significantly influences the two important Raman modes i.e. A$_{1g}$ and E$^1_{2g}$ in MoS$_2$ and WS$_2$ epitaxial thin film grown by PLD. We have observed significant stiffening of two Raman active modes i.e. A$_{1g}$ and E$^1_{2g}$ for MoS$_2$ thin film and mostly A$_{1g}$ for WS$_2$ and the two modes stiffens further with decreasing layer thickness. This translates to a maximum compressive strain of 0.52 % & 0.53 % and increases the band gap to 1.73 eV and 1.68 eV for monolayer MoS$_2$ and WS$_2$, respectively and falls off with increasing the layer thickness. The HRTEM imaging from the interface directly reveals the nature of atomic registry and strain between the films and substrate. The overall results are extremely encouraging and show a practical route to engineer the phonon modes and retain the strain in such van der Waals compounds which can be exploited to tune the opto-electronic properties and fabrication of large area practical devices.

## 2. Experimental Techniques

Thin films of MoS$_2$ and WS$_2$ are grown by pulsed laser deposition (PLD) on '$c$' plane sapphire. We have used 8 sq. mm size of substrate for the present study but little bigger substrate can also be used in our system. MoS$_2$ and WS$_2$ target pellets were prepared from powders obtained from Sigma Aldrich by first cold pressing and then sintering at 500 °C for 5 hours in a vacuum chamber (~$10^{-5}$ Torr). The sintering under large vacuum chambers with continuously running pumps prevents oxidation of compounds as well as re-deposition of vapor species back on the pellet surface compared to sintering performed in a sealed quartz tube. This was a successful method for system containing highly volatile elements [32, 33]. Growth was followed by three step process originally developed for ZnO but with a different temperature and kinetic settings [31]. The temperature for nucleation layer was 400 °C with a laser ablation rate of 1 Hz. The temperature for final growth was 800 °C and at 1Hz laser ablation rate. Both the nucleation layer and slower laser frequency rate are important to obtain epitaxial large area film as we found that faster laser ablation rate leads to formation of polycrystalline $MoS_2$ films [supporting information]. As already explained previously this slow laser ablation rate allows sufficient time for kinetic relaxation of the nucleation layer in order to establish epitaxial relationship with the substrate thus removing misaligned crystallites. This is even more important for van der Waals compounds because the 2D-nucleations do not have any strong attachment with the underlying substrate and is relatively more mobile during the deposition compared to non-layered or three dimensional covalently bonded materials. The method described above was successful for the epitaxial growth of ZnO alloyed with Co, Mn, S and Te on sapphire [31-34]. The pressure was kept constant at ~$10^{-5}$ Torr throughout the growth schedule. There are few recent reports on using sapphire substrate for growing TMDCs films [35-37].

The PLD method of growth has some benefits over chemical vapor deposition (CVD) or physical vapor deposition (PVD) counterparts in the sense that the films will be free from any parasitic deposition. Moreover, the technique is economical and scalable even on a larger substrate than mentioned here in the present study [38, 39].

The epitaxial quality of thin films is confirmed by X-ray diffraction, high resolution transmission electron microscopy and electron diffraction techniques. TEM cross sectional samples were prepared by first mechanical polishing and then Ar ion milling to perforation in order to generate large electron transparent thin area. Special care was taken during the sample preparation as it was found that van der Waals layers can easily be detached from the substrate. The relative movement between two cross sectional pieces was made almost absent to ensure their presence on the substrate till the end of sample preparation.

Raman spectra were recorded using a custom built Raman spectrometer using a 532 nm laser excitation and a grating of 1800 lines/mm at room temperature [40]. The laser power at the sample was approximately 1 mW. Micro PL (photoluminescence) measurement was performed in a state of the art LabRAM HR (UV) system.

## 3. Theoretical Calculation

We have used density functional theory calculation based on WIEN2k code [41] to evaluate the change in electronic structure and band gap of both $MoS_2$ and $WS_2$ monolayer corresponding to the experimentally observed in-plane compressive strain. Generalized gradient approximation (GGA) was performed with Perdew-Burke-Ernzerhof (PBE) functional for optimization of lattice parameters and minimization of forces. The optimized lattice parameters are $a$ = 3.192 & 3.168 Å for $MoS_2$ and $WS_2$ monolayer, respectively. A k-mesh of 19×19×2 for the integration of Brillouin zone and spin orbit coupling was incorporated.

## 4. Results and Discussion

### 4A. Epitaxial growth of $MoS_2$ and $WS_2$ thin films

We have used HRTEM imaging to confirm the formation of epitaxial film, number of layers and stacking information in addition to routinely practiced Raman spectroscopy. On the other hand, X-ray signal does not show up from such one or two layers of films, while spectra from the thick films confirms the formation of large area epitaxial film on sapphire substrate [supporting information]. Hexagonal domains are clearly visible in SEM images shown in supporting information.TEM is extremely powerful technique and which can provide information from 40-50 µm length scale considering four quadrants of the thin areas in a cross sectional TEM specimen. As already mentioned in the TEM sample preparation method, special care must be undertaken during sample preparation in order to retain the weakly held van der Waals layers on the substrate till the end. Fig. 1 display the example HRTEM images with various numbers of layers for $MoS_2$ films and monolayer $WS_2$ film. Both the electron diffraction pattern and FFT of HRTEM images are placed in the inset for the thick films. Diffraction pattern shows <0002> spots parallel to the growth direction. Aberration corrected negative $C_S$ imaging at the interface regions is carried out to probe the nature of layer bonding with the substrate, strain and interlayer stacking information (Fig. 2)

 [42,43]. For the thick film the stacking structure of $MoS_2$ and $WS_2$ are of 2H poly-type. It can be seen that the sapphire substrate is Al terminated and at the end it will be explained that the interaction between the periodic dangling bonds at Al atoms and bottom S layers of $MoS_2/WS_2$ is responsible for the substrate induced compressive strain and associated phonon mode hardening. From the HRTEM images layer specific variation of compressive strain has been estimated averaging over 15 atoms at different regions. The compressive strain is found to be 1.11±0.43 and 0.74±0.33 % for the first and second layers, respectively which are different from the strain derived from the Raman spectra i.e. 0.52 and 0.25 % for monolayer and bilayers, respectively. The possible discrepancy between Raman and HRTEM could be very small area (nanometers) information probed by HRTEM compared to large area (microns) probed by Raman spectroscopy. Strain can also be in-homogenously distributed in the films. Strain fades gradually as one goes away from the film-substrate interface along the '*c*' direction. Extended lattice defects are observed for the thick films and their nature and influence on the physical properties will be the topic of a separate paper.

**4B. Substrate induced layer specific stiffening in $E^1_{2g}$ and $A_{1g}$ Raman modes**

Fig. 3 shows $E^1_{2g}$ and $A_{1g}$ Raman modes of both $MoS_2$ and $WS_2$ for different layer thickness. Hardening of $E^1_{2g}$ and $A_{1g}$ Raman modes indicates the presence of compressive strain [44]. We have observed substrate induced compressive strain in both $MoS_2$ and $WS_2$ layers with the maximum strain present for the monolayer and the strain decreases with increasing the number of layers. Other than stiffening, systematic anomalous shift in the Raman modes are also observed as the number of layers is decreased which is consistent with the earlier reports [26]. Fig. 4 summarizes the shift in $E^1_{2g}$ and $A_{1g}$ Raman modes of $MoS_2$ and $WS_2$ as a function of film thickness grown on 'c' plane sapphire [see supplementary information for tabulated numbers]. The frequency of these two phonon modes for bulk starting powder samples used to grow films in our case and the values from the literature are also listed for the comparison purpose [26]. The Raman shifts of these two modes are slightly different for the powder sample (379 and 404.5 cm$^{-1}$, Sigma Aldrich) used for the present investigation in comparison to the literature values (379.21 and 404.71 cm$^{-1}$) [26]. The difference between the two modes ($\Delta = A_{1g}-E_{2g}$) for the above mentioned two different samples are 25.5 and 24.75 cm$^{-1}$, respectively, and almost close to each other. The difference in absolute values of these two modes between the present powder and the bulk samples may be due to the morphological effect, as powders are composed of particles which are hundreds of microns in size compared to the flat finite size few layer bulk samples reported in the literature. The thicker $MoS_2$ film grown in our case (~ 100 nm or ~ 140-150 number of layers) has the $E^1_{2g}$ and $A_{1g}$ values as 383.37 and 410.13 cm$^{-1}$ with $\Delta = 26.76$. The absolute values are almost close to the literature reported multilayer values except slightly larger $\Delta$ may be because of large number of layers over large areas (at least 8×8 sq. mm) compared to the previous reports (Fig..3). This possibly introduces stronger long range Coulomb force compared to the micrometer sized bulk powder and softens the $E^1_{2g}$ mode further and consequently increases the band gap. The frequency values shifts to the higher wave numbers or stiffens with decreasing the number of layers as seen for the Raman shift of E and A modes in $MoS_2$ with shifts for monolayer/bilayer being 391.77/387.32 and 409.88/409.32 cm$^{-1}$ (Fig. 3(a)). The difference between the Raman shifts of the two modes, $\Delta$ is 18.11 and 21.9 compared to the reported values of 17.18 and 21.4 cm$^{-1}$, respectively for monolayer and bilayer [26]. Thus Raman spectroscopy also confirms the formation of monolayer and bilayers MoS$_2$ larger area epitaxial thin film in the present case in addition to TEM imaging. We also observe anomaly in the shifts in two Raman modes for MoS$_2$ films on sapphire substrate with number of layers as reported earlier except E$^1_{2g}$ mode in our case stiffens significantly from 383.37 in the thick film to 387.32 and 391.77 cm$^{-1}$ for the bilayers and monolayer, respectively whereas A$_{1g}$ mode softens from 410.13 for the thick film to 409.88 for the monolayer. For monolayer, the difference in the two Raman modes between literature and the present case is 3.92 and 2.49 for the E$^1_{2g}$ and A$_{1g}$ modes, respectively [26]. This difference in Raman modes translates into equivalent compressive strain of 0.52 % in the monolayer film on sapphire substrate. To best of our knowledge this is the only report on the substrate induced strain and its retention and this is important for tuning of optoelectronic property of such material in an effective and efficient way. It has already been predicted that introducing compressive stress will increase the band gap of the system. The change in band gap is expected to be 1.73 eV (1.68 eV without strain in PBE-GGA) for the monolayer MoS$_2$ (Fig. 4). The corresponding indirect band gap also increases from 1.88 eV to 1.93 eV. The origin of strain based on HRTEM imaging is discussed in the subsequent section.

PL spectra shows clear emission at 1.98 eV from monolayer film but no clear signal is obtained from bilayer and thick films suggesting that monolayer property is retained on sapphire substrate but crossover takes place after depositing subsequent layers on top of monolayer (Fig. 5).

For WS$_2$ it is already reported that the A$_{1g}$ (Γ) mode stiffens with increasing the number of layers which is similar to the behavior of MoS$_2$. This stiffening of A$_{1g}$ mode is due to the increase in restoring force caused by band renormalization through interlayer Coulomb coupling and van der Waals interaction [45]. On the other hand the E$^1_{2g}$ phonon mode shows only subtle changes with the number of layers. The bulk WS$_2$ micron size powders used in our case shows frequency corresponding to E$^1_{2g}$ and A$_{1g}$ as 352.63 and 420.5 cm$^{-1}$, respectively and the values reported in the literature are 355.5 and 420.5 cm$^{-1}$ [46] with Δ being 67.87 and 65 cm$^{-1}$, respectively. WS$_2$ thin films grown on sapphire substrate do not show any significant changes in the frequency of $A_{1g}$ mode, but $E^1_{2g}$ mode is observed to harden significantly on decreasing the number of layers with the shifts being 363.5 and 362.56 cm$^{-1}$ with $\Delta$ = 55 and 55.92 for monolayer and bilayer, respectively. This is significantly different than reported values for monolayer i.e. 355.9 and 417.5 cm$^{-1}$, with a frequency difference 61.6 cm$^{-1}$. This shows the significant influence of sapphire substrate on the $WS_2$ layer. This difference in Raman modes translates into an equivalent compressive strain of 0.53 % in the monolayer film on sapphire substrate. The band gaps of monolayer $WS_2$ are 1.68 for direct. However, with 0.53 % compressive strain it is at the edge of cross over from direct (1.682 eV) to indirect (1.678 eV) band gaps (Fig. 4). Clear PL emission peak at 1.97 eV is obtained in the case of monolayer $WS_2$ as well and the state of the material at the edge of crossover did not disturb the probability of emission across the direct band gap. The situation is different from the case of $MoS_2$ where crossover is expected to take place at a compressive strain of ~ 2% [47]. This is because the rate of change of both types of band gaps with compressive strain is different for $WS_2$ and $MoS_2$. The band gap modification in both the cases can be understood in terms of coupling between various S and Mo valence orbitals [47].

Few reports exist on the stiffening of the Raman modes and associated increase in compressive strain [22, 27, 29]. In previous reports, strain was applied by external means. But in our case the strain is induced by substrate and can be retained which will be useful for practical device engineering of this material. The origin and nature of this substrate induced strain is discussed next.

**4C. Nature of substrate induced strain and interlayer stacking**

It is interesting that though both $MoS_2$ and $WS_2$ are van der Waals compounds, sapphire substrate is capable of inducing compressive strain in this material. Generally, van der Waals interaction is comparatively very weak compared to usual ionic or covalent chemical bonding and for this force it may seem difficult to sustain the strain. However, the theoretical calculation for 0.52% compressive strain corresponding to monolayer $MoS_2$ shows that the strain energy cost is only 10 meV which is much lower compared to interlayer van der Waals force of 460 meV for $MoS_2$. This increases the band gap to 1.74 eV from 1.68 eV obtained using PBE-GGA potential based calculation and a value of 1.98 eV obtained by PL measurement. This implies that, there must be weak interaction existing between $MoS_2$ and the underlying substrate. Negative $C_s$ HRTEM imaging directly reveals the interfacial structure of both $MoS_2$ and sapphire. We present only results based on $MoS_2$ and similar result can be expected for $WS_2$. Fig. 2 shows the atomic resolution interface image of $MoS_2$ on '*c*' plane sapphire. $MoS_2$ has following orientation relationship with sapphire substrate i.e. <11-20>$Al_2O_3$ || <01-10> $MoS_2$. The schematic model is shown in Fig. 2 and a strain of 1.11±0.43 and 0.74±0.33 % can be derived for the first and second layer, respectively from the image. Also, from the HRTEM imaging it is found that this interaction also translates to the third layer and beyond and reduces with the thickness of the film. The interlayer stacking is found to be Bernal stacking (2H poly-type, Fig. 2). Earlier reports mentioned about wide range of misoriented domains of $MoS_2$ grown epitaxially on a substrate and this was explained based on subtle difference in energies between different orientations [35, 36]. However, in our case both by HRTEM and X-ray only one type of oriented grains are observed probably due to the slower kinetics employed [31]. Nevertheless, question remains regarding the role of substrate and its polarity on the band gap of such material. In this context, it was already shown theoretically that O dangling bond affects most among various dangling bonds in $MoS_2$ thin films on $SiO_2$ substrate [48]. O dangling bonds reduces the indirect band gap for both the monolayer and multi layers films significantly compared to direct band gap but for Si- terminated surface or H-passivated surface, changes are subtle and direct band gap remains fundamental. Thus, as it is clear from the HRTEM imaging that in the present case the substrate is Al- terminated and it is the interaction between Al dangling bonds and S which introduces the strain in these films. Based on the recent theoretical results in Ref. 49, Al terminated sapphire substrate influences the electronic structure of $MoS_2$ only weakly. The experimental Al-S distance is found to be ~2.7 Å which is almost close to 2.6 Å predicted by theory [49].

The theoretical calculation also shows that $MoS_2$ can grow on sapphire with different degree of misorientation due to small energy difference between them. From our HRTEM imaging at the interface it is clearly visible that pairing of two S atoms around the Al atoms due to strain (Fig. 2). Therefore, in the presence of compressive strain the difference between the direct and indirect gaps will increase as expected and remain suitable for practical device exploration and application.

## 5. Conclusion

In summary, we have grown large area epitaxial thin films of van der Waals compounds $MoS_2$ and $WS_2$ on '$c$' plane sapphire by pulsed laser deposition with control over the number of layers. Substrates induced compressive strain is responsible for the stiffening of both the Raman modes with implication in increase in the direct band gap of these materials and their retention. The results demonstrate a practical way to engineer the optoelectronic property of such materials for large area device fabrication and application.

**Supporting Information**.

HRTEM images of Polycrystalline of $MoS_2$, XRD of epitaxial thin film, Raman spectrum of bulk powder, Table of Raman data

**Notes**
The authors declare no competing financial interest.

**ACKNOWLEDGMENT**
The authors at JNCASR sincerely acknowledge Prof. C.N.R. Rao for providing advanced microscopy and crystal growth facility for this research. D.S.N. and B.V. thanks CSIR, India for funding.

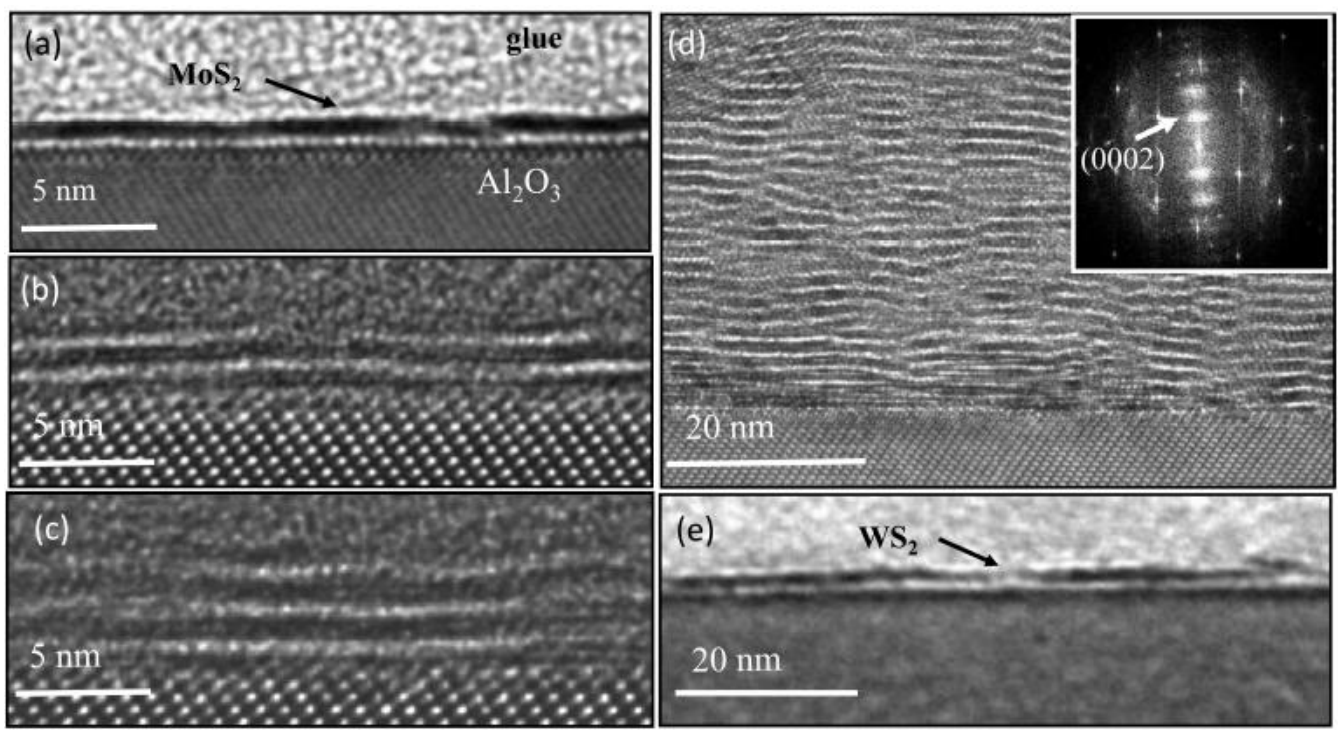

**Fig. 1.** HRTEM images showing (a) monolayer, (b) bilayers, (c) few layers, and (d) thick epitaxial films of MoS$_2$ on '*c*' plane sapphire. Epitaxial monolayer WS$_2$ on '*c*' plane sapphire is shown in (e). Both the electron diffraction and X-ray spectra confirm the formation of large area epitaxial thin film on sapphire. X-ray spectra can be found in the supplementary document.

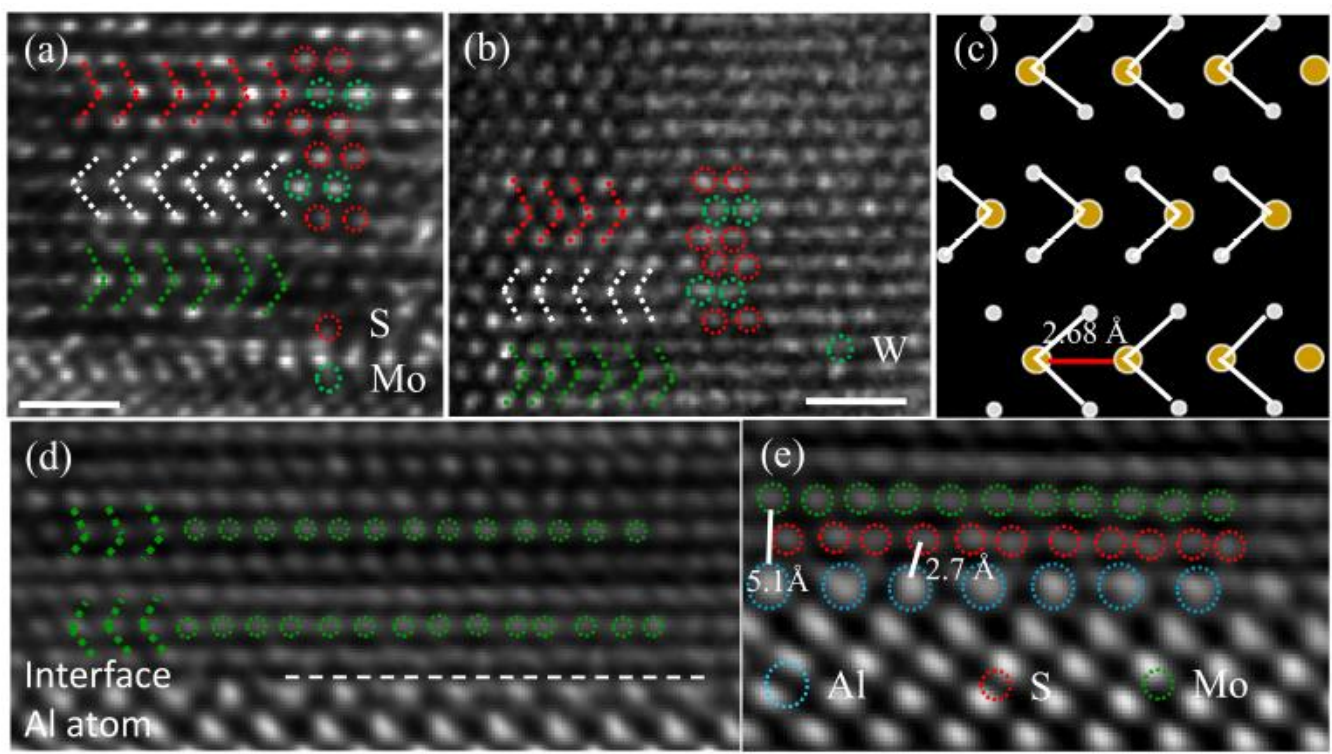

**Fig. 2.** HRTEM imaging from the interface regions of epitaxial $MoS_2$ and $WS_2$ films along <11-20> Z.A. of sapphire. (a) & (b) showing the 2H poly-type interlayer stacking for $MoS_2$ and $WS_2$, respectively. (d) Example image from $MoS_2$/sapphire interface markings atoms over a distance from place to place to derive strain information from such images. (e) Close up interface image showing that substrate is Al terminated and the typical distance between Al and S is ~ 2.7 Å. A modulated in plane S-S (marked with red dotted circle) inter-atomic distance over Al (marked turquoise dotted circle) atoms can be observed.

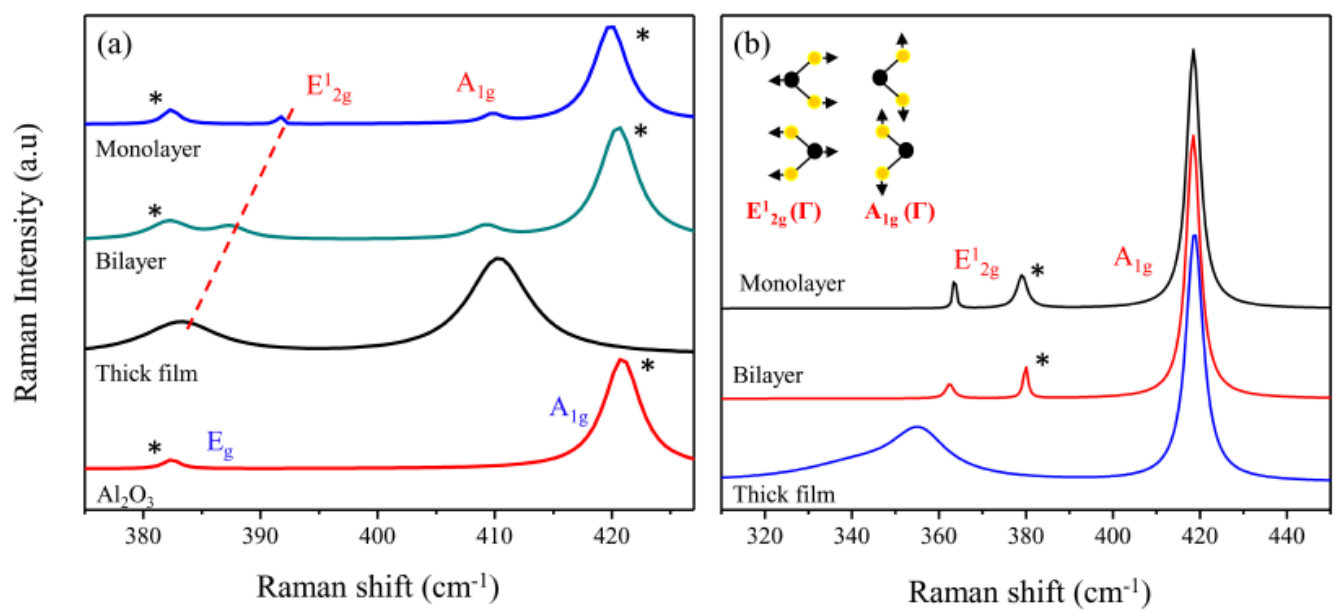

**Fig. 3.** The $A_{1g}$ and $E^1_{2g}$ Raman modes for monolayer, bilayer and thick films for (a) $MoS_2$, and (b) $WS_2$. For comparison the Raman modes of bare sapphire is also shown in (a). Raman shift for bulk powder from literature can be found in the supplementary information. The inset of (b) shows the atomic displacements of the Raman modes. Peaks marked with the asterisks are from sapphire substrate.

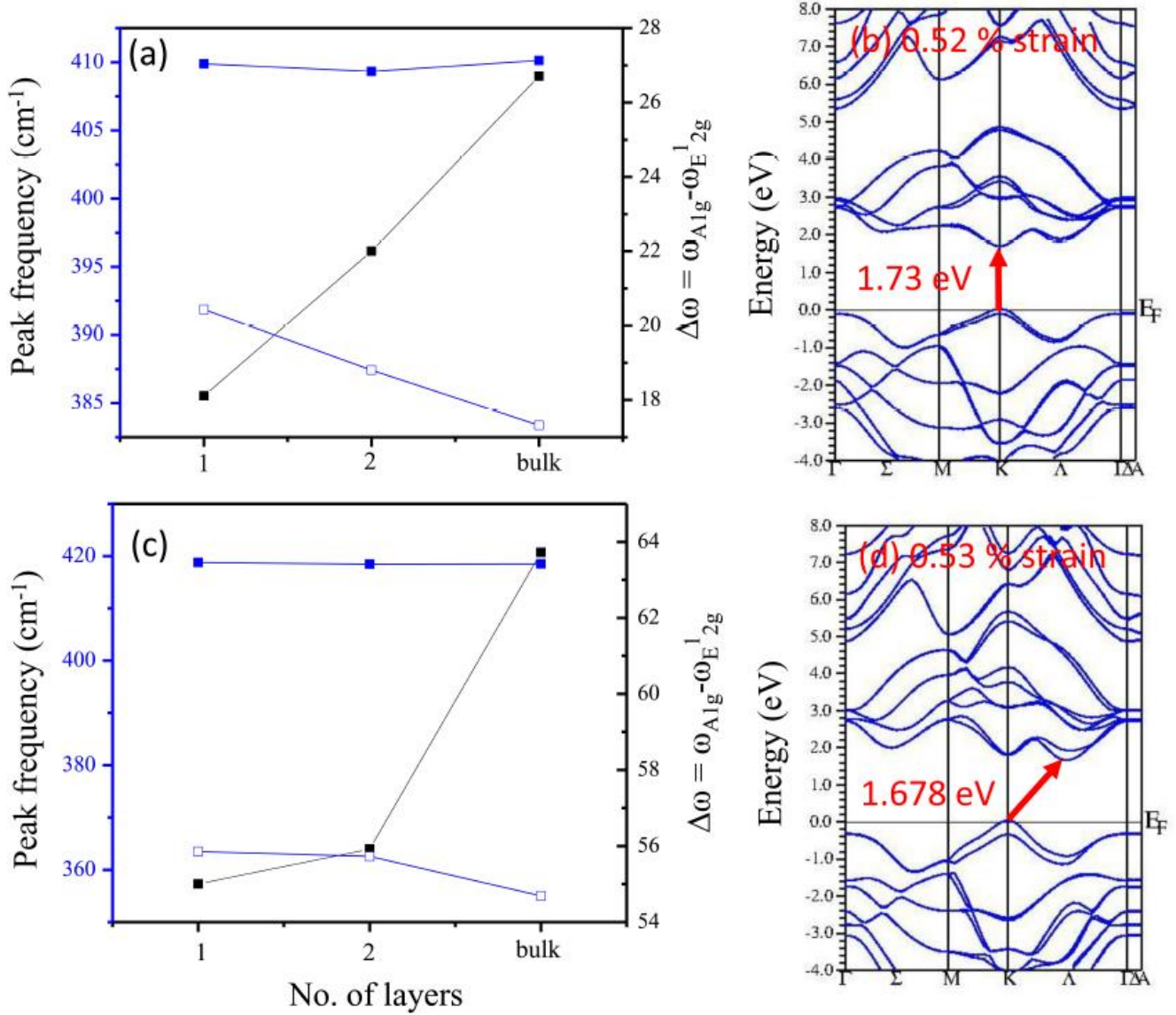

**Fig. 4.** Raman shifts of $A_{1g}$ and $E^1_{2g}$ modes for monolayer, bilayers and thick films for (a) $MoS_2$ and (b) $WS_2$, showing stiffening due to compressive strain with decreasing number of layers. The difference between the $A_{1g}$ and $E^1_{2g}$ peaks showing anomalous peak shift with decreasing number of layers is also plotted (blue). (c) & (d) are the calculated band structure corresponding to the experimentally observed strain. The direct band gap of $MoS_2$ increases and remains fundamental, but for similar magnitude of compressive strain $WS_2$ shows emergence of cross over from direct to indirect band gap.

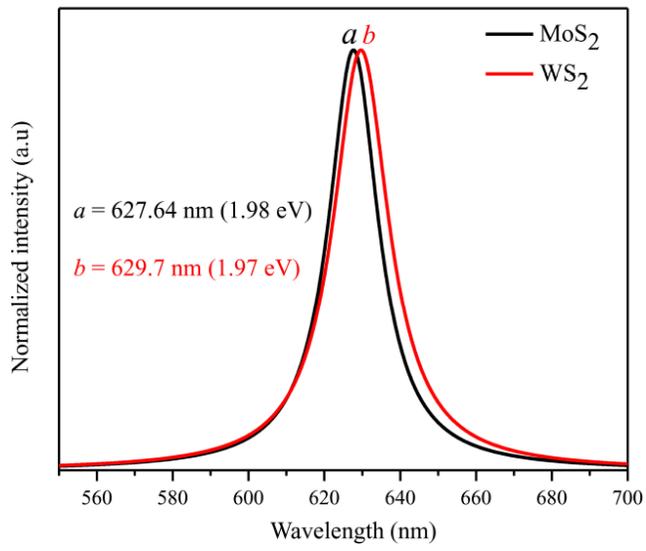

**Fig. 5.** Photoluminescence spectra of monolayer MoS$_2$ and WS$_2$ epitaxial thin film on '*c*' plane sapphire.

# Substrate induced tuning of compressive strain and phonon modes in large area MoS$_2$ and WS$_2$ van der Waals epitaxial thin films

Rajib Sahu[1], Dhanya Radhakrishnan[2], Badri Vishal[1], Devendra Singh Negi[1], Anomitra. Sil[3], Chandrabhas Narayana[2], and Ranjan Datta[*,1]

[1]International Centre for Materials Science, Jawaharlal Nehru Centre for Advanced Scientific Research, Bangalore 560064, India.

[2] Light Scattering Laboratory, Chemistry and Physics of Materials Unit, Jawaharlal Nehru Centre for Advanced Scientific Research, Bangalore 560064, India.

[3]Center for Nano Science and Engineering, Indian Institute of Science, Bangalore 560012, India.

*Corresponding author E-mail:* [ranjan@jncasr.ac.in](ranjan@jncasr.ac.in)

SEM images :

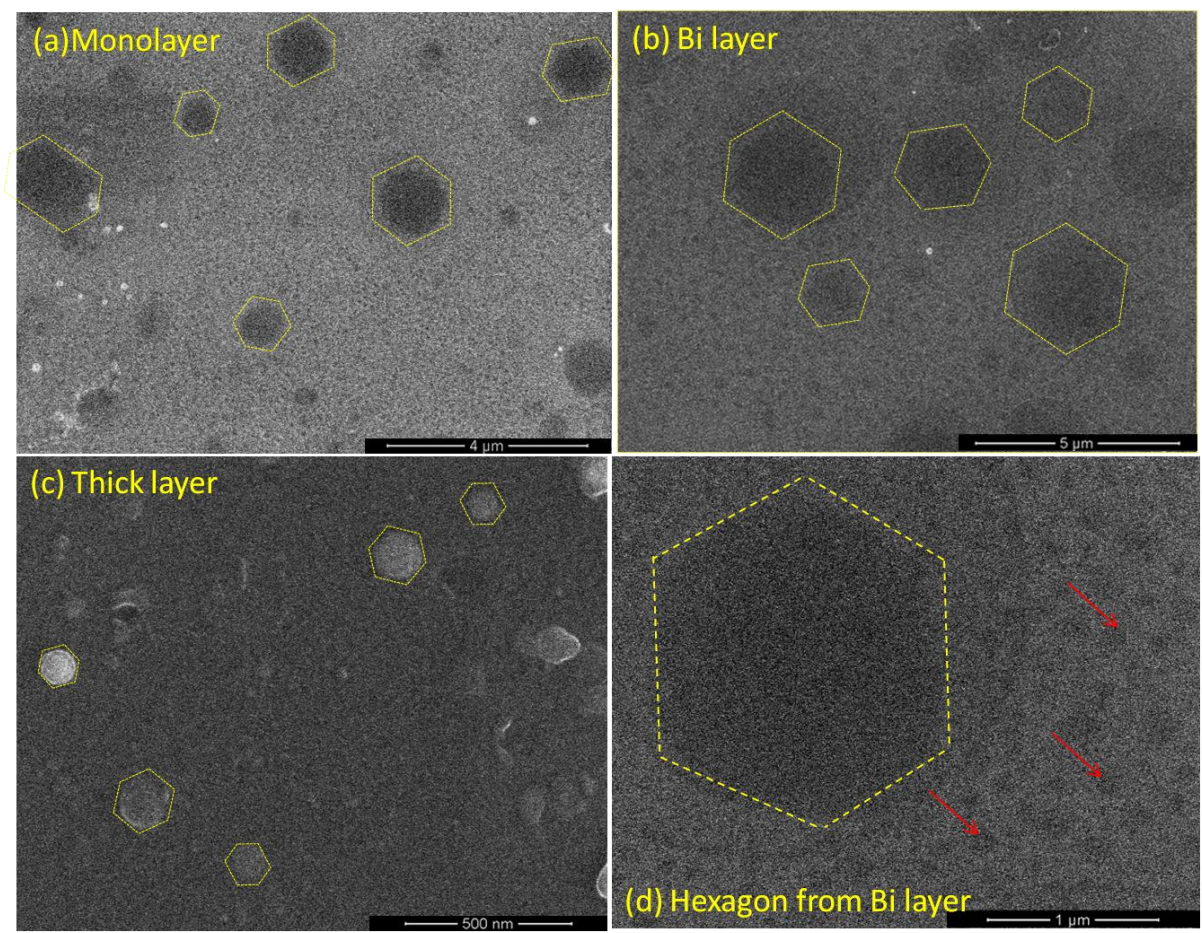

**Fig. S1**. SEM images of (a) monolayer, (b) & (d) bi-layer & (c) thick film. Hexagonal domains are clearly visible from all three samples. The larger hexagons from mono and bi-layer films are due to large area crystallites and smaller dark areas as marked with the red arrows in (d) are the small size crystallites. The coverage is extremely good to intercept these during cross sectional TEM imaging.

TEM imaging:

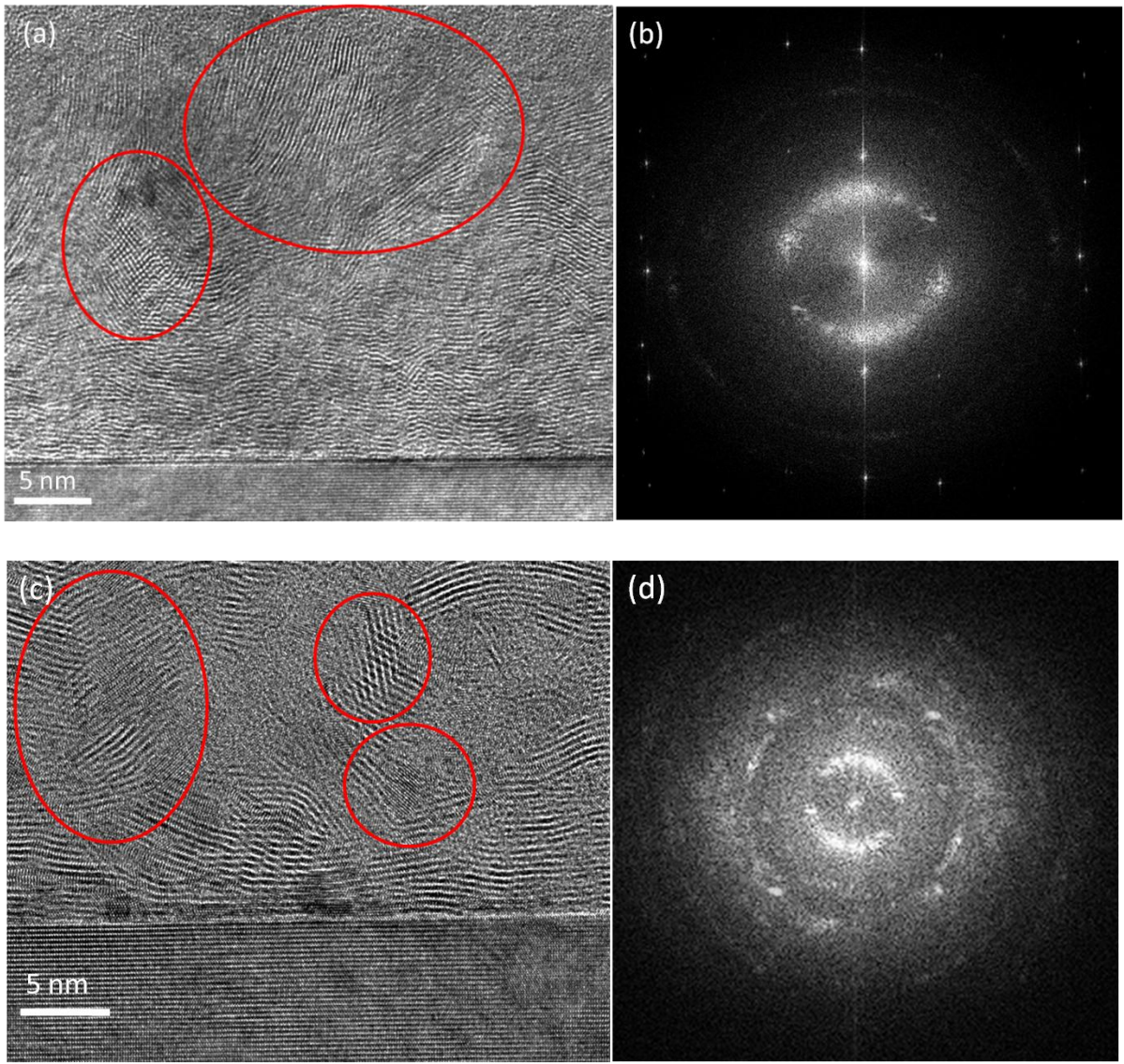

**Fig. S2**. TEM images showing formation of polycrystalline films under faster kinetic condition. Formation of epitaxial film can only be ensured under slow kinetic growth for the initial buffer layer.

XRD :

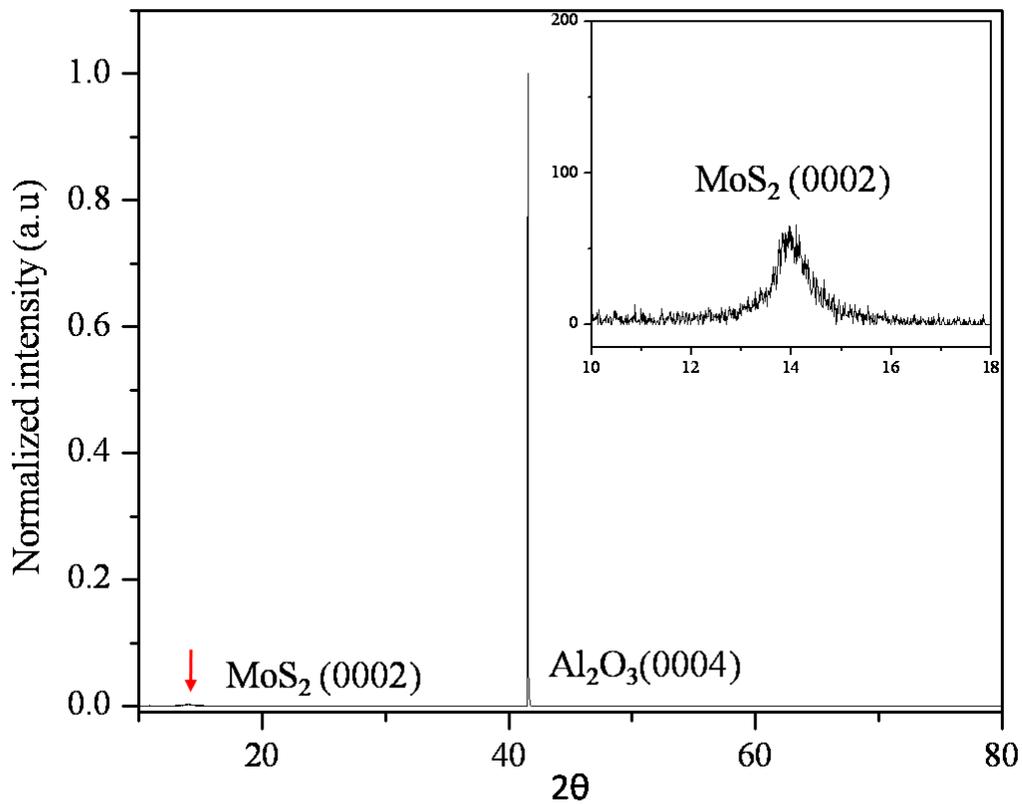

**Fig. S3**. X-ray spectra confirming the formation of large area epitaxial $MoS_2$ thin film on '$c$' plane sapphire. 0002 and 0004 peaks corresponding to $MoS_2$ and $Al_2O_3$ are marked. For mono layer and bilayers it was not possible to obtain a clear signal by X-ray. The reason for this is that for coherent Bragg X-ray diffraction at least few numbers of layers is required to obtain any out of plane peaks.

Raman spectra:

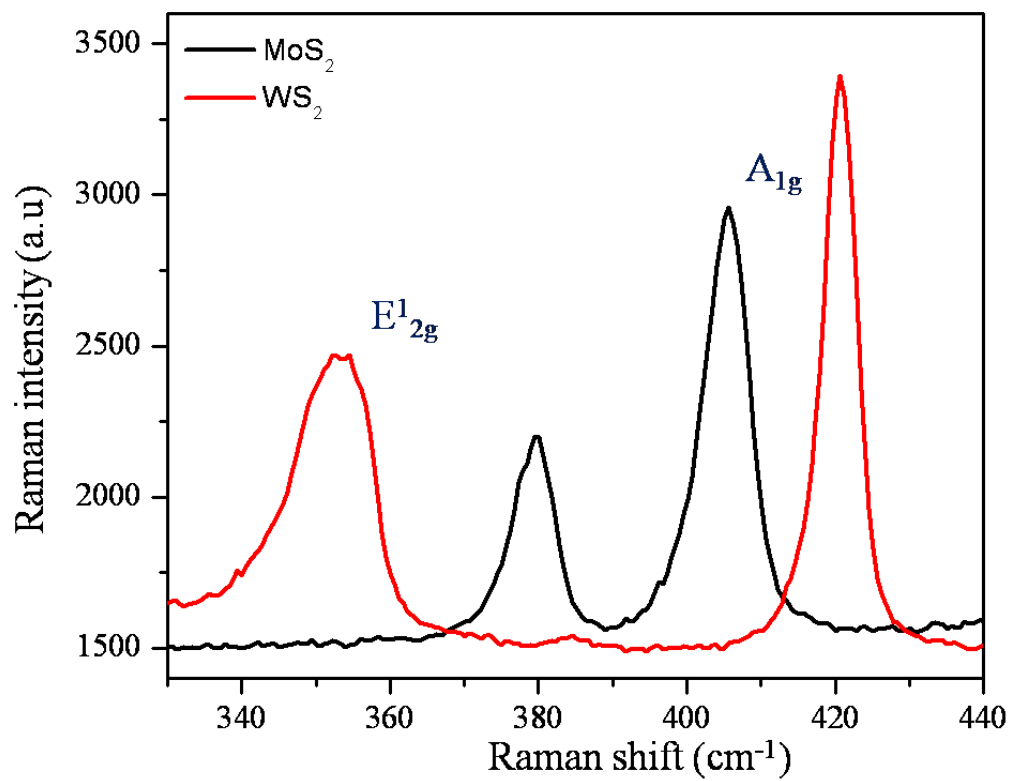

**Fig. S4**. Raman spectra showing the $E^1_{2g}$ and $A_{1g}$ modes at 379.93, and 353.6 cm$^{-1}$, 405.71 and 420.7 cm$^{-1}$ for MoS$_2$ and WS$_2$ powder samples, respectively.

**Table 1:** Raman shift for the characteristic $A_{1g}$ and $E^1_{2g}$ Raman modes of $MoS_2$.

| $MoS_2$ | E12g | A1g | Δ | Comp.strain w.r.t literature | Comp.strain w.r.t thick film | Band gap (eV) |
|---|---|---|---|---|---|---|
| Thick Film | 383.37 | 410.13 | 26.71 | | | |
| Bi layer | 387.42 | 409.32 | 22 | 0.27% | 0.28% | 1.4 |
| Mono layer | 391.77 | 409.88 | 18.11 | 0.51% | 0.52% | 1.736 |
| Literature (bulk) | 383.48 | 408.23 | | 24.75 | | |

**Table 2:** Raman shifts for the characteristic $A_{1g}$ and $E^{1}_{2g}$ Raman modes of $WS_2$.

| $WS_2$ | $E12g$ | $A1g$ | $\Delta$ | Comp.strain w.r.t literature | Comp.strain w.r.t thick film | Band gap (eV) |
|---|---|---|---|---|---|---|
| Thick Film | 355.04 | 418.76 | 63.72 | | | |
| Bi layer | 362.56 | 418.48 | 55.92 | 0.44% | 0.47% | |
| Mono layer | 363.39 | 418.5 | 56.11 | 0.49% | 0.52% | 1.678 |
| Literature (bulk) | 355.5 | 420.5 | 65 | | | |